\documentclass[aps,prl,showpacs,twocolumn,superscriptaddress]{revtex4-1}
\usepackage{bm,color,amsmath,amssymb,mathrsfs,latexsym,graphicx,psfrag}

\begin{document}
\title{Functional renormalization group for the anisotropic triangular antiferromagnet} 

\author{Johannes Reuther} \affiliation{Institut~f\"ur~Theorie~der~Kondensierten~Materie,~Karlsruhe Institute of Technology,~76128 Karlsruhe,~Germany} 

\author{Ronny Thomale}
\affiliation{Department of Physics, Princeton University, Princeton,
  NJ 08544, USA} 

\begin{abstract}
  We present a functional renormalization group scheme that allows us
  to calculate frustrated magnetic systems of arbitrary lattice
  geometry beyond $\mathcal{O}(200)$ sites from first principles. We
  study the magnetic susceptibility of the antiferromagnetic (AFM)
  spin-$1/2$ Heisenberg model ground state on the spatially anisotropic triangular lattice,
  where $J'$ denotes the coupling strength of the intrachain bonds
  along one lattice direction and $J$ the coupling strength of the
  interchain bonds. We identify three distinct phases of the
  Heisenberg model.  Increasing $\xi=J'/J$ from the effective square
  lattice $\xi=0$, we find an AFM N\'eel order to spiral order
  transition at $\xi_{c1} \sim 0.6-0.7$, with indication to be of second
  order. In addition, above the isotropic point at $\xi_{c2} \sim
  1.1$, we find a first order transition to a magnetically disordered
  phase with collinear AFM stripe fluctuations.
\end{abstract}
\date{\today}


\maketitle

Frustration induced by lattice geometry and interaction in
two-dimensional quantum antiferromagnets constitutes a considerable
challenge beyond mean field theory~\cite{misguichlhuillier04}. This is
the main reason why the question of distinguishing systems with
magnetic long-range order and spin liquid behavior has been one of the
most difficult and long-standing problems in the field of quantum
magnetism. A plethora of different methods has been developed to
address this problem in the context of many different lattices and
magnetic Hamiltonians, most of which, however, have severe drawbacks
of very different kind. Quantum Monte Carlo (QMC)
methods~\cite{sandvik-91prb5950} may suffer from the sign problem
often encountered for frustrated systems, while exact diagonalization
(ED) studies are constrained to small system sizes.  Density matrix
renormalization group (DMRG)
methods~\cite{white92prl2863,schollwock05rmp259} partly resolve the
latter problem, but are constrained to (effectively) one-dimensional
systems.  Coupled cluster methods (CCM)~\cite{farnellbishop04} and
variational QMC methods~\cite{capriotti-99prl3899} are valuable
approaches to treat frustrated two-dimensional systems. However, they
are limited in the sense that, generally, only trial state energies
can be tested against each other, and no unbiased treatment from first
principles is possible.

Promising lines of improvement have been undertaken recently. In
special cases where it is applicable, dimer projection schemes allow
to treat larger systems at a similar level of precision and
completeness as ED~\cite{poilblanc-09cm0724}. From a perturbative
expansion perspective, continuous unitary transformation (CUT) methods
provide a helpful tool to compute the energies of the ground state
and excitation modes of small quasiparticle
sectors~\cite{wegnercut,milaschmidt}. Furthermore, the generalized
notion of
matrix product states, in certain cases, leads to an efficient
treatment of two-dimensional magnetic systems by projected entangled
pair states (PEPS)~\cite{verstraete-08adp143}. In particular, the
multi-scale entanglement renormalization Ansatz (MERA) provides a new
tool to compute energies and scaling behavior of certain magnetic
systems, and has been most recently applied to frustrated
systems~\cite{evenbly-09prl180406,evenbly-09cm3383}. Still, even in
the optimal cases where the approximations made within these methods
are controlled, it is mainly suited to determine certain ground state
properties only.
For frustrated magnetic systems and actual comparison to experiment,
however, it would be most desirable to compute complementary
thermodynamic quantities that allow to resolve the competition and
classification of magnetic ordering and quantum fluctuations. The most
suitable quantity in this respect is the magnetic susceptibility.
Interpreted as the magnon spectral function, it provides detailed
information about the qualitative and quantitative type of magnetic
fluctuations, and is the canonic quantity measured in experiment.  In
this Letter, we employ the pseudofermion functional
renormalization group (PFFRG)~\cite{reuther-10prb144410} as a new
method to tackle systems of frustrated magnetism.  Applying the method
to the anisotropic triangular lattice, we demonstrate that the PFFRG
is able to (i) treat large system sizes of $\mathcal{O}(200)$ sites,
(ii) is applicable to arbitrary frustrated lattice geometries and
two-body bare interactions, (iii) naturally allows to compute the magnetic
susceptibility as the canonical outcome of the RG, and (iv) hence
provides an unbiased calculation from first principles that allows
comparison to experiment.

\begin{figure}[t]
\begin{minipage}[l]{0.99\linewidth}
\includegraphics[width=\linewidth]{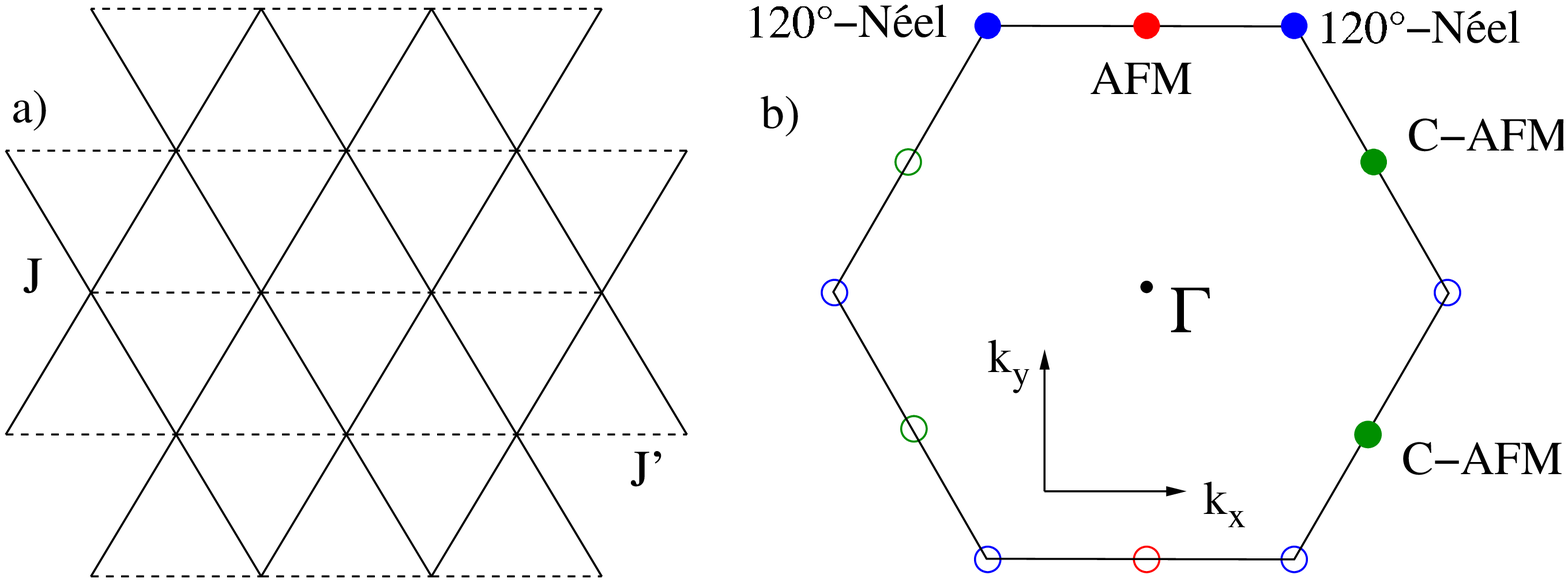}
\end{minipage}
\caption{(color online) a) Triangular lattice structure. The
  horizontal bonds correspond to coupling $J'$ the other to coupling
  $J$. b) Schematic plot of the hexagonal Brillouin zone. Different
  magnetic order resides at different points (shown: AFM N\'eel order,
  $120^{\circ}$ N\'eel order, collinear AFM order (c-AFM)). The open
  circles relate to the filled circles by reciprocal lattice vectors,
  i.e., AFM order corresponds to two points in the BZ, collinear AFM
  to four points, and $120^{\circ}$ N\'eel order to six points.}
\label{pic1}
\vspace{-0pt}
\end{figure}
The Heisenberg model on the anisotropic triangular lattice (ATLHM)
attracted considerable attention in recent years as an experimentally
accessible testing ground for quantum magnetism disorder phenomena.
The Hamiltonian is given as
\begin{equation}
H_{\text{ATLHM}}=J \sum_{\langle i,j \rangle_{\textnormal{v}}} \vec{S}_i \vec{S}_j + J' \sum_{\langle i,j \rangle_{-}} \vec{S}_i \vec{S}_j, 
\label{atlhm}
\end{equation}
where the coupling $J'$ applies to the bonds along (horizontal)
one-dimensional chains and $J$ is the coupling between them, forming a
triangular lattice altogether (Fig.~\ref{pic1}a). We define $\xi=J'/J$
as a parameter to interpolate between the effective square lattice
limit $\xi=0$ and the disordered isolated chain limit $\xi \rightarrow
\infty$.  Experiments on
$\text{Cs}_2\text{CuCl}_4$~\cite{coldea-01prl1335} ($\xi \sim 2.94$)
provide a good testing ground of discussing various features of
spin liquid behavior. The formation of a
magnetically ordered state for smaller $\xi$ opposed to disorder
tendencies for larger $\xi$ can be nicely studied for the organic
$\kappa-(\text{BEDT}-\text{TTF})_2\text{X}$ family. While
$\text{X}=\text{Cu}_2[\text{N}(\text{CN})_2]\text{Cl}$ shows an AFM
transition of $T_\text{N}=27 \text{K}$ with estimated $\xi=0.55$, the
$\text{X}=\text{Cu}_2(\text{CN})_3$ compound, estimated to be nearby
the symmetric triangular regime $\xi=1.15$, does not show magnetic
order down to very low temperatures~\cite{shimizu-03prl107001}.
Details of the phase diagram of~\eqref{atlhm} are still of current
debate. It is an established fact that the system is AFM
N\'eel-ordered for small $\xi$, changing to $120^{\circ}$ N\'eel order as the
isotropic triangular limit $\xi=1$ is
reached~\cite{heidarian-09prb012404,kohno-09prl197203,weng-06prb012407,merion-99jcm2965,shen-02prb172407,bishop-09prb174405,pardini-08prb214433}.
However, an intermediate disordered phase
has been proposed~\cite{weihong-99prb14367}, while
other works assume a direct transition, but cannot classify the
transition to be of second or first
order~\cite{bishop-09prb174405,shen-02prb172407}. For larger $\xi > 1
$, some works claim a disordered phase extending to the $\xi
\rightarrow \infty$ limit~\cite{weng-06prb012407,yunoki-06prb014408},
whereas others indicate collinear antiferromagnetic (c-AFM)
ordering~\cite{bishop-09prb174405,pardini-08prb214433,starykh-07prl077205}.
 
\begin{figure}[t]
\begin{minipage}[l]{0.99\linewidth}
\includegraphics[width=\linewidth]{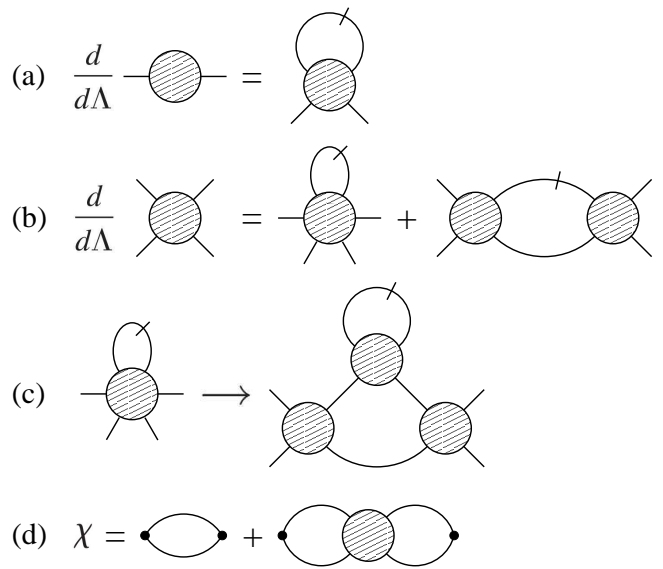}
\end{minipage}
\caption{Graphic representation of the FRG differential equations. Bare
  lines denote the (renormalized and scale dependent) Green's
  functions and slashed lines the single scale propagators. Eqs. (a)
  and (b) are the FRG equations for the self-energy and the
  two-particle vertex, respectively (without distinguishing between
  the different pairing channels on the right side of (b)). The
  Katanin scheme is given by the replacement (c). Fusing the external
  legs of the two-particle vertex, the magnetic susceptibility is
  obtained as shown in (d).}
\label{pic2}
\vspace{-0pt}
\end{figure}
\begin{figure*}[t]
\begin{minipage}{0.99\linewidth}
\includegraphics[width=\linewidth]{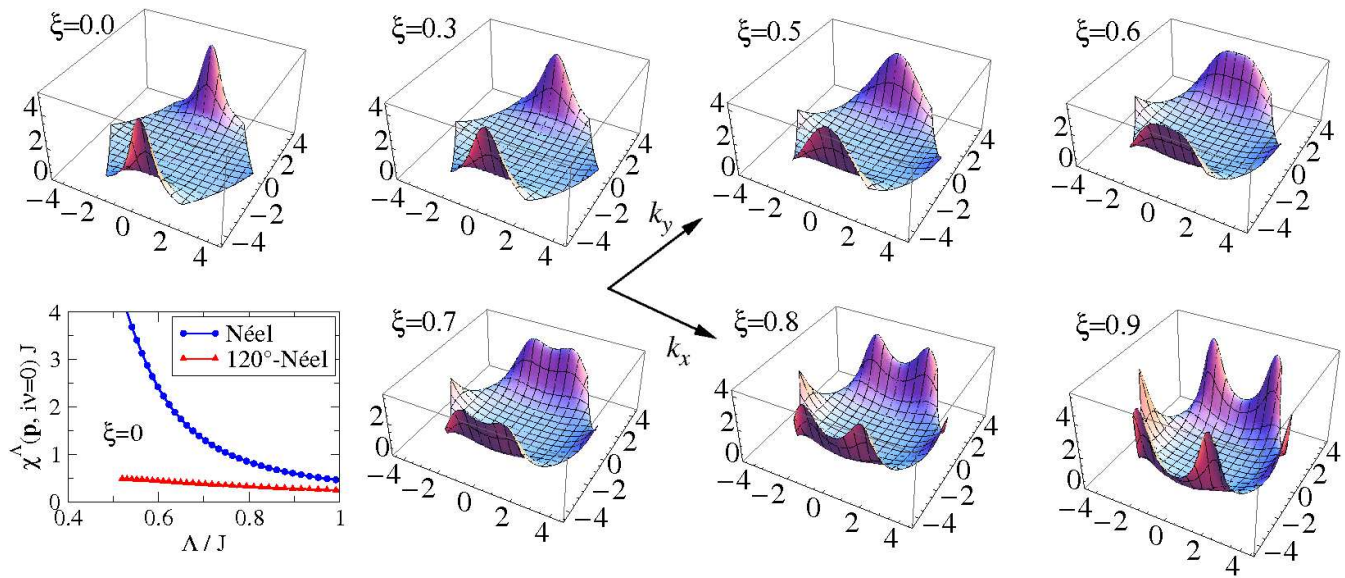}
\end{minipage}
\caption{(color online) Static magnetic susceptibility resolved for
the whole
  Brillouin zone, varying $\xi=J'/J$ from the effective square lattice
  $\xi=0.0$ close to the triangular lattice $\xi=0.9$. Bottom Left:
  2-particle vertex flow for the AFM N\'eel channel (blue) versus the
  120$^{\circ}$ N\'eel channel (red) at $\xi=0$ with respect to the IR
  cutoff flow parameter $\Lambda$. We observe that the AFM vertices start to
  diverge, signalling a magnetic instability.  Susceptibilities are
  always given in units of $1/J$. The respective types of order for
  the distinct peaks positions are shown in Fig.~\ref{pic1}b.}
\label{pic3}
\end{figure*}
\begin{figure*}[t]
\begin{minipage}{0.99\linewidth}
\includegraphics[width=\linewidth]{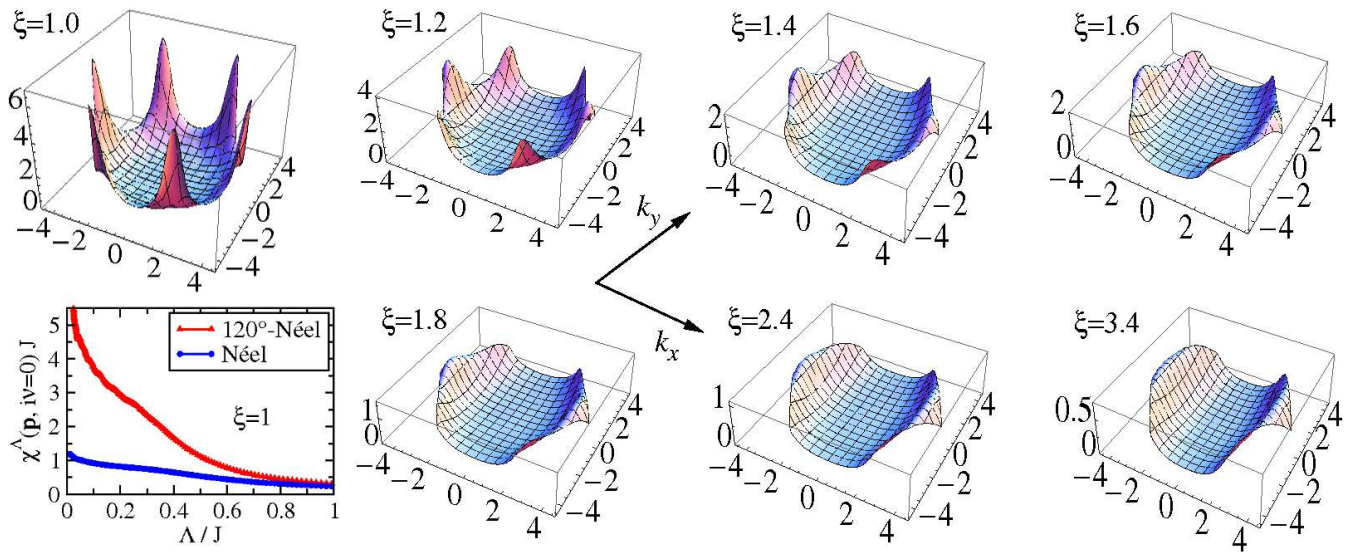}
\end{minipage}
\caption{(color online) Static magnetic susceptibility, varying $\xi$
  from the isotropic triangular lattice $\xi=1.0$ towards the 1d chain
  limit $\xi \rightarrow \infty$. Bottom left: Vertex flow at the
  $\xi=1$ point. The rise of the $120^{\circ}$ N\'eel channel shows
  the ordering instability at the isotropic triangle point. Compared
  to the flow at $\xi=0$ (Fig.~\ref{pic3}), the rise takes place at a
  much lower scale of $\Lambda$, indicating a lower ordering scale for
  the triangular lattice.}
\label{pic4}
\end{figure*}

We now address this problem with PFFRG, which is explained in more
detail in~\cite{reuther-10prb144410}. Unlike conventional FRG studies,
our starting point is not given by the bare excitations of the spin system.
Instead, we use the 
pseudofermion representation of spin-1/2 operators $S^{\mu} = 1/2
\sum_{\alpha\beta} f_{\alpha}^{\dagger} \sigma_{\alpha\beta}^{\mu}
f_{\beta}$, ($\alpha,\beta = \uparrow,\downarrow$, $\mu = x,y,z$) with
the fermionic operators $f_{\uparrow}$ and $f_{\downarrow}$ and the
Pauli-matrices $\sigma^{\mu}$. This representation enables us to apply
Wick's theorem leading to
standard Feynman many-body techniques. The introduction of
pseudofermions comes along with an artificial enlargement of the
Hilbert space and therefore requires the fulfillment of an occupancy
constraint (exclusion of empty and doubly occupied states). Since an
unphysical occupation acts as a vacancy in the spin lattice associated
with an excitation energy of order $J$, particle number fluctuations
are suppressed at zero temperature and the constraint is naturally
fulfilled. Quantum spin models are inherently strong coupling models,
requiring infinite self-consistent resummations of perturbation theory. In this
context
FRG~\cite{wetterich93plb90,morris-94ijmpa2411,honerkamp-01prb035109,hedden-04jpcm5279}
provides a systematic summation in different interaction channels by
generating equations for the evolution of all one-particle irreducible
$m$-particle vertex functions under the flow of an IR frequency cutoff
$\Lambda$ (see Fig.~\ref{pic2}a and 2b for the flow of the self-energy
and the two-particle vertex). In order to reduce the infinite
hierarchy of equations to a closed set, some truncation scheme has to be applied. As an important difference in the
PFFRG as compared to conventional FRG schemes, we still include certain 2-loop contributions that are shown by
Katanin~\cite{katanin04prb115109} to be essential for a better
fulfillment of Ward-identities (see Fig.~\ref{pic2}c). In particular,
this way the random phase approximation (RPA) is recovered as a diagram
subset generated by the RG flow~\cite{salmhofer-04ptp943}. It is
important to emphasize that the RPA diagrams play a crucial role in
our scheme as they are responsible for the collective ordering
phenomena. 
In addition, however, our method treats the leading diagrammatic contributions of
a $1/S$ expansion~\cite{anderson52pr694,brinkmann-04prb174445} (direct
particle-hole graphs) , $1/N$
expansion~\cite{affleck-88prb3774,marston-89prb11538} (crossed
particle-hole graphs), and particle-particle diagrams on the same footing. This enables
us to adequately account for the competition of magnetic order and disorder
in an unbiased fashion. The magnetic susceptibility can be
conveniently computed from the two-particle vertex (Fig.~\ref{pic2}d).
Due to the local nature of the auxiliary fermions of a bare spin
model, a real-space representation of all vertices is much more
suitable than the usual momentum space scheme. Furthermore, since we
operate in the strong coupling limit, the proper regularization of
Green's functions requires the self-consistent back coupling of the
self energy (Fig.~\ref{pic2}a) into the propagators in
Fig.~\ref{pic2}b. In order to account for the dynamic fluctuations, it
is also necessary to keep all frequency dependencies of the vertex
functions involved. The numerical solution requires the discretization
of these frequencies. Similarly, the spatial dependence is
approximated by keeping correlation functions of an infinite system up
to a maximal length.  In our calculations, this length extends over 7
lattice spacings, leading to a correlation area of 169 lattice sites
(this constrains the resolvable incommensurable order to vector sizes
within that range). An ordering instability is initially signalled by
a strong rise of the vertex couplings associated with this order at
some finite scale of $\Lambda$ (Fig.~\ref{pic3} and~\ref{pic4}). Since
in the present formulation rotational invariance is conserved during
the flow, we should not find stable solutions down to $\Lambda=0$ in
magnetically ordered regimes. Indeed, the onset of spontaneous long
range order is signalled by a sudden breakdown of the smooth flow. In
contrast, the existence of a stable solution indicates the absence of
long range order associated with spin rotational symmetry breaking. In
the present implementation, our approach does not directly detect
chiral order. This is because the chirality operator is a three-body
operator by construction, and hence relates to the 3-particle vertex
whose explicit RG flow is not computed.

We sweep through the parameter space of $\xi$ from the square lattice
to the isotropic triangular lattice and compute the static magnetic
susceptibility shown in Fig.~\ref{pic3} (the peak positions for
different types of long range order are depicted in Fig.~\ref{pic1}b).
Throughout this parameter regime, we observe a characteristic
breakdown of the flow, indicating ordering instabilities rather than a
disordered phase. One can nicely observe how the susceptibility
evolves as we increase $\xi$. 
As shown in Fig.~\ref{pic3}, we find a broadening and subsequent
splitting of the N\'eel-peak. The splitted peaks evolves along the Brillouin-zone edge and the weight at the corners of the hexagon increases until in the isotropic case the peak symmetrically resides at the corner positions. From the splitting of the N\'eel peak, we estimate the
transition to be at $\xi_{c1} \sim 0.6-0.7$, i.e., the regime above which
the order becomes incommensurate.  
Apparently, the system influenced
by quantum fluctuation favors AFM fluctuations over spiral
fluctuations, since the classical transition point at $\xi=0.5$ is
shifted to higher $\xi$. In particular, from the extremely smooth
evolution of the susceptibility upon variation of $\xi$ which we
studied by steps of $0.02$ in $\xi$, we find
indication for the transition to be of second order, while an
extremely weak first order transition (corresponding to a slight kink
in the leading susceptibility channel) cannot be excluded as a matter
of principle. The height of the peaks such as in Fig.~\ref{pic3} still
depend on system size. At the isotropic triangular point
$\xi=1$ where additional lattice symmetries enable us to consider
systems even beyond 250 sites, we obtain higher and sharper peaks with
increasing system size.

In a recent work, Starykh and Balents have found that for $\xi > 1$,
the quantum system enters a collinear antiferromagnetic stripe
phase~\cite{starykh-07prl077205}. This is interesting since the
classical estimate would be spiral order in that regime, so the
quantum fluctuations lead to a different
ordering~\cite{pardini-08prb214433}. However, finite size numerical
studies reported a disordered phase in that
regime~\cite{shen-02prb172407}. Our results for the magnetic
susceptibility for $\xi \geq 1$ are shown in Fig.~\ref{pic4}. As
depicted, we observe a strong drop in the magnetic susceptibility
above the isotropic point, i.e., in the regime $\xi_{c2} \gtrsim
1.1$. From here, no ordering instability is found in the RG flow and
the susceptibility rapidly looses the $120^{\circ}$ N\'eel order
signature. While the $120^{\circ}$ N\'eel order peaks die out quickly,
AFM stripe fluctuation signatures emerge (at points of the Brillouin
zone according to Fig.~\ref{pic1}). The transition appears to be of
first order according to a pronounced kink in the maximal
susceptibility upon varying $\xi$.  While we do not find a breakdown
of the flow that would indicate magnetic ordering, we still obtain
strong collinear AFM stripe fluctuations (in agreement
with~\cite{starykh-07prl077205}) signalled by an unstable RG flow that
develops oscillations sensitively depending on the frequency
discretization. These fluctuation tendencies are also seen at higher
$\xi$ where the peak structure is still visible along the
$k_{x}$-direction. However, these peaks are strongly broadened along
the $k_{y}$-axis, i.e., smeared between the two c-AFM ordering vector
positions. This indicates a fast exponential decay of spin
correlations between the $J'$-chains. There are proposals in the
literature that the (supposedly) disordered regime splits into two
(gapped or gapless) different spin liquid
phases~\cite{yunoki-06prb014408}. In principle this would correspond
to a change of diffuse spectral weight in the magnetic susceptibility
going from an ungapped to a gapped system. We do not find clear
indication for this scenario. At large $\xi$ the susceptibility does
not change significantly with increasing $\xi$, which we studied by
computing scenarios up to $\xi=5$ in steps of $0.2$ in $\xi$.

In conclusion, we have used the pseudo-fermion functional
renormalization group to study the different phases of the anisotropic
triangular lattice. We find that upon variation of the anisotropy
parameter $\xi$, the system divides into a
N\'eel order, spiral order, and a disordered phase with c-AFM
stripe fluctuations. We find evidence for a second order transition
between the first two and a first order transition between the last
two of these phases. 
We are  confident that our method is
a suitable starting point to discuss various other problems in the
field of frustrated magnetism.

\begin{acknowledgments}
  We thank P.~Schmitteckert for his numerical support.  
  We thank D.~A.~Abanin, B.~A.~Bernevig, A.~L\"auchli, K.~P.~Schmidt, 
  S.~Sondhi, O.~Tchernyshyov, S.~Wessel, and, in particular,
  P.~W\"olfle for discussions. JR is supported by DFG-FOR 960. RT is
  supported by a Feodor Lynen Fellowship of the Humboldt Foundation
  and Alfred P. Sloan Foundation funds.
\end{acknowledgments}


{\bf Supplementary material}

\begin{figure*}[t]
\begin{minipage}[l]{0.99\linewidth}
\includegraphics[width=\linewidth]{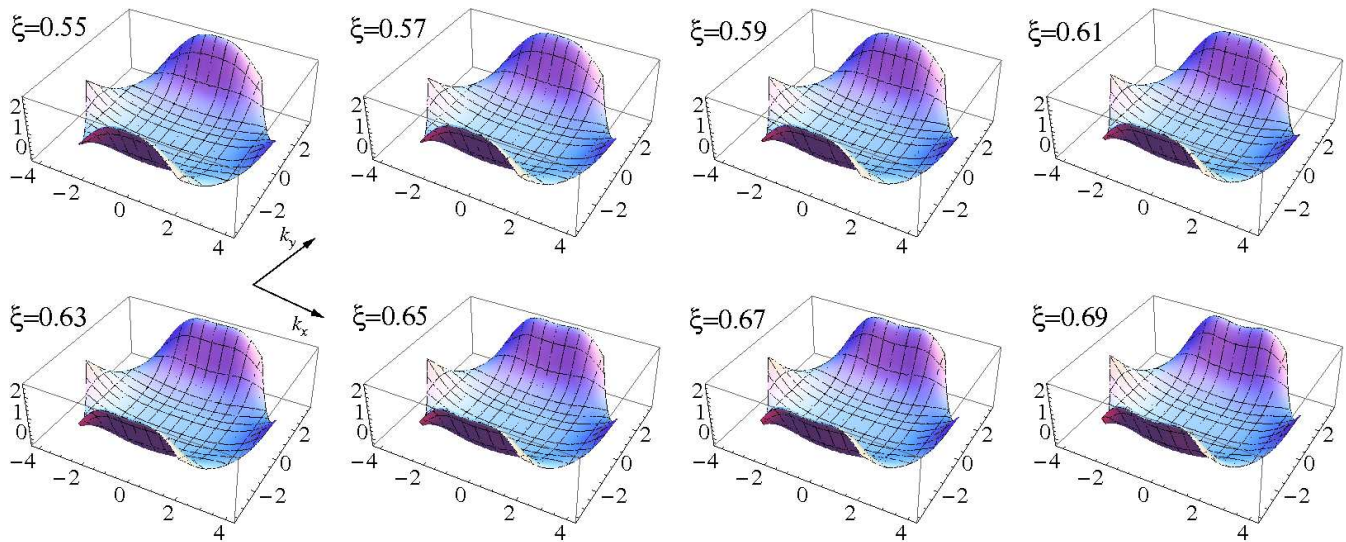}
\end{minipage}
\caption{Static magnetic susceptibility, varying $\xi$
  from the AFM N\'eel side across the phase transition to
  spiral ordering. The already broadened AFM N\'eel peak splits up
  into two incommensurate peaks that successively shift towards the
  edges of the Brillouin zone where the $120^{\circ}$ N\'eel order
  peaks reside.}
\label{pic5}
\end{figure*}

{\it N\'eel order to spiral order transition.}  In the following we
have resolved the N\'eel order to spiral order transition in higher
resolution varying the anisotropy parameter $\xi$ in small stepwidths
(see Fig.~\ref{pic5}). We find that the previous N\'eel peak first
broadens along $k_x$ and then smoothly splits up into two peaks which
then evolve along the Brillouin-zone boundary. This process comes
along with increasing spectral weight at the corners of the Brillouin
zone. Note that as a consequence of the periodicity in momentum space,
the emerging peak structure at $k_y=0$ presents the tails of the
broadened N\'eel peak. Increasing $\xi$ further towards the isotropic
triangular point (see Fig.~\ref{pic3}) the split peaks move towards
the corner position until at $\xi=1$ the hexagonal symmetry of the
susceptibility is reached. As the susceptibility evolves completely
smoothly through the transition, we find it to be of second order. We
identify the wave vector of the corresponding long range-ordered
phases with the position of the maximal susceptibility. From this we
locate the transition point at such $\xi$ where the peaks split and
above which the order becomes incommensurate. From Fig.~\ref{pic5} a
transition at $\xi\approx0.61$ can be read off. In the calculations
for this figure we kept correlations up to a length of 5 lattice
spacings (as compared to 7 lattice spacings in Fig.~\ref{pic3} and
Fig.~\ref{pic4}). In addition, the number of discrete frequencies has been
slightly reduced in Fig.~\ref{pic5}. These reductions generally result
in lower and broader peaks, which demonstrates that (according to the
fact that the phases in this parameter regime are long range-ordered)
there are still finite size effects in the susceptibility. Especially
at the isotropic triangular point where additional lattice symmetries
enable us to treat much longer correlations, we are able to resolve
peaks at bigger system sizes which, in units of Fig.~\ref{pic4}, reach heights up to 12
in $1/J$, while the structure of the susceptibility is unchanged. However, in order to be able to detect long range order, for
which the breakdown of the smooth RG-flow is the relevant criterion,
we found it sufficient to use 5 or 7 lattice spacings as the maximal
length of correlations. Another effect seen in a finite size scaling
is that with increasing length of correlations (and also increasing
number of discrete frequencies), the transition point moves towards slightly
higher $\xi$. Therefore from a finite size scaling we estimate the
transition to be approximately centered between $\xi=0.6$ and
$\xi=0.7$.
\begin{figure*}[t]
\begin{minipage}[l]{0.99\linewidth}
\includegraphics[width=\linewidth]{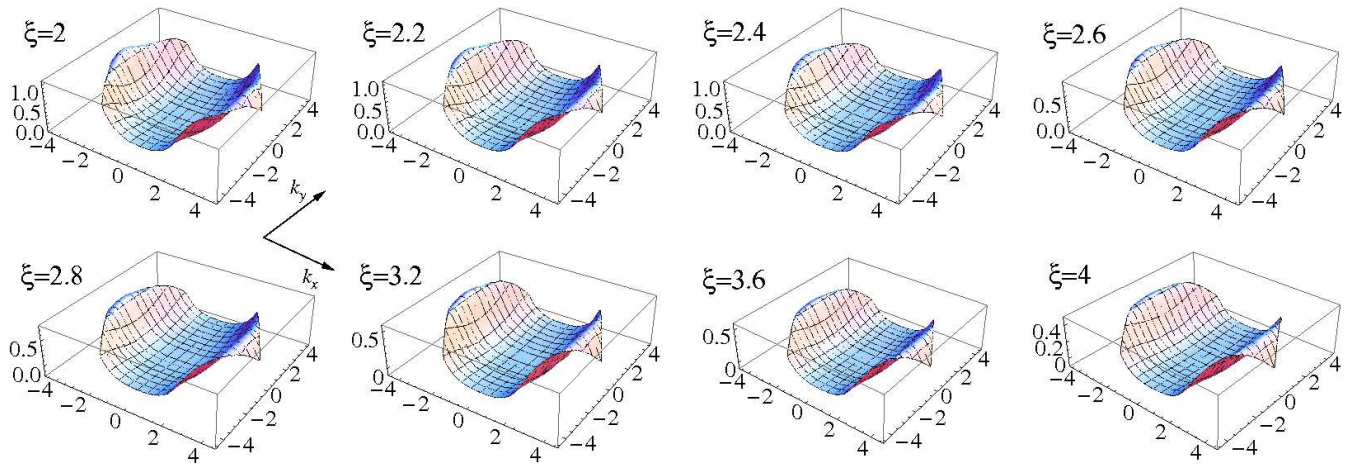}
\end{minipage}
\caption{Static magnetic susceptibility, varying $\xi$
  in the strongly anisotropic regime from $\xi=2$ to $\xi=4$. While
  the spectral structure remains rather invariant of C-AFM type which
  is strongly smeared along the $k_y$ direction, the total spectral
  weight slightly decreases with increasing $\xi$.}
\label{pic6}
\vspace{-0pt}
\end{figure*}

{\it High anisotropy regime.}  We provide more detailed data on the
strong anisotropy regime (Fig.~\ref{pic6}). As found there, the strong
decay of correlations between the effectively weakly coupled
one-dimensional chains is already dominant, which manifests itself in
a strong smearing of the magnetic susceptibility along $k_y$. As $\xi$
is increased, the spectral weight of the susceptibility gets more and
more reduced, while the fluctuation structure remains rather
stable. We only observe reminiscences of C-AFM fluctuations smeared
along $k_y$. From there, we do not find clear indication for a bi-spin
liquid scenario suggested in the literature for that regime, where one
magnetically disordered phase should be gapped while the other should
be gapless [Yunoki and Sorella, Phys. Rev. B 74, 014408 (2006).]. A
transition between those phases should manifest itself in the
susceptibility by some jump in the spectral weight. In general, at
large anisotropies, energy scales involved in the coupling between the
chains become so small that a caveat has to be given from the
frequency discretization incorporated in the numerics, which sets a
lower bound of energy scales to be resolved by our method. We varied
the frequency mesh in the lower energy regime, and found no notable
variation of our results. While this should hold for the anisotropy regime
shown in the manuscript, we cannot exclude different physical
scenarios for even higher anisotropy.

\end{document}